\documentclass[10pt,useAMS,usenatbib]{mn2e}
\usepackage{mn2ejour}
\usepackage{amssymb}
\usepackage{color}
\usepackage{bm}
\usepackage{graphicx}

\title[Vortex stretching]{
Vortex stretching in self-gravitating protoplanetary discs
}

\author[Zs. Reg\'aly and E. Vorobyov]{Zs. Reg\'aly$^{1}$\thanks{E-mail:regaly@konkoly.hu} and E. Vorobyov$^{2,3}$\\
$^1$Konkoly Observatory, Research Centre for Astronomy and Earth Sciences, Hungarian Academy of Sciences,\\
\,\,\,\,1121, Budapest, Konkoly Thege Mikl\'os \'ut 15-17, Hungary\\
$^2$Department of Astrophysics, University of Vienna, 1180, Vienna, Austria \\
$^3$Research Institute of Physics, Southern Federal University, Stachki Ave. 194, 344090, Rostov-on-Don, Russia
}

\begin{document}


\pagerange{\pageref{firstpage}--\pageref{lastpage}} \pubyear{2017}
 
\maketitle
\label{firstpage}

\begin{abstract}
Horseshoe-shaped brightness asymmetries of several transitional discs are  thought to be caused by large-scale vortices. Anticyclonic vortices are efficiently collect dust particles, therefore they can play a major role in planet formation. Former studies suggest that the disc self-gravity weakens vortices formed at the edge of the gap opened by a massive planet in discs whose masses are in the range of $0.01\leq M_\mathrm{disc}/M_*\leq0.1$. Here we present an investigation on the long-term evolution of the large-scale vortices formed at the viscosity transition of the discs' dead zone outer edge by means of two-dimensional hydrodynamic simulations taking disc self-gravity into account. We perform a numerical study of low mass, $0.001\leq M_\mathrm{disc}/M_*\leq 0.01$, discs, for which cases disc self-gravity was previously neglected.  The large-scale vortices are found to be stretched due to disc self-gravity even for low-mass discs with $M_\mathrm{disc}/M_*\gtrsim0.005$ where initially the Toomre $Q$-parameter was  $\lesssim50$ at the vortex distance. As a result of stretching, the vortex aspect ratio increases and a weaker azimuthal density contrast develops. The strength of the vortex stretching is proportional to the disc mass. The vortex stretching can be explained by a combined action of a non-vanishing gravitational torque caused by the vortex, and the Keplerian shear of the disc. Self-gravitating vortices are subject to significantly faster decay than non-self-gravitating ones. We found that vortices developed at sharp viscosity transitions of self-gravitating discs can be described by a GNG model as long as the disc viscosity is low, i.e. $\alpha_\mathrm{dz}\leq10^{-5}$.
\end{abstract}

\begin{keywords}
accretion, accretion discs --- hydrodynamics --- instabilities --- methods: numerical --- protoplanetary discs  
\end{keywords}

\section{Introduction}

Dozens of transitional discs show remarkable brightness asymmetries (see e.g. \citealp{Brownetal2009, Andrewsetal2009, Hughesetal2009, Isellaetal2010,Andrewsetal2011, Mathewsetal2012, Tangetal2012, Fukagawaetal2013, Casassusetal2013,vanderMareletal2013,Perezetal2014,Hashimotoetal2015,Casassusetal2015,Wrightetal2015,Momoseetal2015,Marinoetal2015}.  The horseshoe-shaped asymmetries seen in the millimetre-wavelength images are thought to be caused by large-scale anticyclonic vortices, although other phenomena, such as disc eccentricity excited by a massive companion star in binaries \citep{Ragusaetal2017} or  self-shadowing caused by a tilted inner disc with respect to the orbit of an inclined giant planet \citep{DemidovaGrinin2014}, can also lead to the development of horseshoe-shaped features.

Vortex formation can be triggered by the baroclinic instability \citep{KlahrBodenheimer2003,LyraKlahr2011,Raettigetal2013,Lyra2014} or by the Rossby wave instability (RWI, first described by \citealp{Rossbyetal1939}) via the coagulation of smaller scale vortices \citep{Lovelaceetal1999, Lietal2000, Lietal2001}. The RWI is excited at the vortensity minimum, which can develop at the pressure bumps in protoplanetary discs, e.g., at the edges of a gap opened by an embedded planet  \citep{Lietal2005}, at the edges of the accretionally inactive zone of discs \citep{VarniereTagger2006,Lyraetal2009b,Meheutetal2010, Crespeetal2011,Meheutetal2012a, Meheutetal2012b, Meheutetal2012c, Regalyetal2012,Meheutetal2013, Richardetal2013,Flocketal2015}, or in the outer regions of protostellar discs accreting from natal clouds \citep{Baeetal2015}. 

Since the pressure maximum formed at the eye of anticyclonic vortices is capable of collecting dust \citep{AdamsWatkins1995,BargeSommeria1995,Tanga1996,KlahrHenning1997,Braccoetal1999,GodonLivio2000}, large-scale vortices can play a crucial role in planet formation if they are a long-lasting phenomenon  (see e.g., \citealp{KlahrBodenheimer2006,HangKenyon2010,OwenKollmeier2017}). However, the disc viscosity is known to reduce the strength and the lifetime of the vortices formed at gaps opened by massive planets \citep{deValBorroetal2007,Ataieeetal2013,Fuetal2014b,Mirandaetal2016}. At the same time, the main source of disc viscosity -- the magnetorotational instability (MRI) -- has recently been brought under question with new non-ideal magnetohydrodynamics simulations indicating that, e.g.,  ambipolar diffusion may act to reduce the strength of the MRI \citep{Bai2013,Gressel2015}. 

Vortices can also be destroyed by the dust accumulated in the vortex via the the dust feedback, if the local dust-to-gas mass ratio approaches unity \citep{Johansenetal2004,InabaBarge2006,Lyraetal2009a,Fuetal2014a,Crnkovic-Rubsamenetal2015,Survilleetal2016}. In the shearing box simulations, vortices are found to be only transient structures due to the effect of disc self-gravity, inhibiting the formation of a single large-scale vortex \citep{MamatsashviliRice2009}.  However, the dust back-reaction on to the gas that causes the vortex decay is found to be only a temporary effect and vortices can be re-established \citep{Raettigetal2015}. Recently, \citet{Mirandaetal2017} investigated the dust feedback on large-scale vortices formed at viscosity transitions and found that it does not inhibit, but rather slows down the process of azimuthal dust trapping. 

In this paper, we consider disc self-gravity as another physical phenomenon that can reduce the strength and lifetime of the vortices, but unlike the MRI, does not depend on the local conditions in the disc, such as temperature, ionization fraction, or shear. Here, we speak about a general long-range action of gravity and not about the development of gravitational instability, which does depend on the local conditions in the disc. Disc self-gravity is known to delay or even suppress the formation of large-scale vortices developed at the planetary gap edges for sufficiently high-mass discs, as was shown by \citet{LinPapaloizou2011} in two-dimensional and \citet{Lin2012} in three-dimensional models. \citet{Baeetal2015} found that large-scale vortices developed in the outer
regions of protostellar discs subject to mass-loading from natal clouds dissipate as the Toomre $Q$-parameter \citep{Toomre1964} reaches unity. citet{ZhuBaruteau2016} have shown that disc self-gravity weakens the vortex development at artificial density bumps.  \citet{LovelaceHohlfeld2013} and recently \citet{Yellin-BergovoyUmurhan2016}, however, have shown that the disc self-gravity should be included in the RWI simulations for discs with the Toomre parameter $Q<Q_\mathrm{crit}=(1/h)$, where $h$ is the geometric aspect ratio (here $G=1$ code unit is applied, see details in the next section). For a flat-disc-surface geometry with $h=0.05$ this corresponds to $Q\mathrm{crit}=20$. Assuming a canonical protoplanetary disc mass (e.g., $M_\mathrm{disc}/M*\simeq0.01--0.1$), size (e.g., $R_\mathrm{out}\simeq30$\,au), and surface mass density slope (e.g., $\Sigma\sim R^{-1}$) the Toomre parameter at a distance where the vortices form in our models ($R\simeq20$\,au) is in the range of $30--300$ suggesting that disc self-gravity has no significant effect on RWI. However, prior to the excitation of the RWI, a density jump forms at the viscosity transition where the Toomre parameter can be decreased to a value where disc self-gravity becomes important.

Here, we present an investigation of the large-scale vortex development at viscosity transitions in protoplanetary discs. We study the effect of disc self-gravity on both the formation and long-term evolution of vortices by means of two-dimensional (thin-disc) hydrodynamic simulations assuming gravitationally stable discs with $0.001\leq M_\mathrm{disc}/M_*\leq0.01$. We show that disc self-gravity stretches the large-scale vortices, making them azimuthally elongated and shortening their lifetimes for relatively low mass discs whose mass is $M_\mathrm{disc}/M_*\gtrsim0.005$. Thus, the effect of disc self-gravity is found to be essential even for relatively low-mass discs.

The outline of this paper is as follows. In Section~2, we describe our two-dimensional numerical hydrodynamics model, which takes the effect of disc self-gravity into account. In Section~3, we present our findings regarding the effect of disc self-gravity on the formation and the long-term evolution of large-scale vortices. In Section~4, we discuss our results and provide an explanation for the vortex stretching. The paper closes with our conclusions and outlooks in Section~5.

\section{Numerical model}

We investigate the formation and evolution of a large-scale vortex developed at a sharp viscosity transition by means of two-dimensional hydrodynamical simulations. In order to model the long-term evolution of vortices we investigate the vortex formation only at the outer dead zone edge. For this investigation we use the GPU supported version of the {\small FARGO} code \citep{Masset2000}, which numerically solves the vertically integrated continuity and Navier--Stokes equations:

\begin{equation}
\frac{\partial \Sigma}{\partial t}+\nabla \cdot (\Sigma {\bf v})=0,
\label{eq:cont}
\end{equation}
\begin{equation}
\frac{\partial {\bf v}}{\partial t}+({\bm v \cdot \nabla}){\bf v}=-\frac{1}{\Sigma} \nabla P+\nu \Delta {\bm v}-\nabla \Phi_\mathrm{tot},
\label{eq:NS}
\end{equation}
where $\Sigma$ and ${\bm v}$ are the two-dimensional surface mass density and velocity vector of the gas.  We use the $\alpha$ prescription for the disc viscosity, in which case $\nu=\alpha c_\mathrm{s}H$, where $c_\mathrm{s}$ is the sound speed and $H$ is  the local pressure scale height \citep{ShakuraSunyaev1973}. The total gravitational potential of the disc in Equation~(\ref{eq:NS}) is
\begin{equation}
\Phi_\mathrm{tot}(r,\phi)=-G\frac{M_*}{r}+\Phi_\mathrm{ind}(r,\phi)+\Phi_\mathrm{sg}(r,\phi),
\label{eq:phi_tot}
\end{equation} 
where the first term is the gravitational potential of the star in a given cell with radial distance $r$. Since the equations are solved in the cylindrical coordinate system centred on the star, Equation~(\ref{eq:phi_tot}) includes the so-called indirect potential, $\Phi_\mathrm{ind}(r,\phi)$ arising due to the displacement of the barycentre of the system caused by any disc non-axisymmetry (see its importance in, e.g., \citealp{MittalChiang2015,ZhuBaruteau2016, RegalyVorobyov2017}). The indirect potential is calculated as
\begin{equation}
\Phi_\mathrm{ind}(r,\phi) =  r\cdot G\int { 
\mathrm{d}m(\bm{r}^\prime) \over {\bm r}^{\prime 3} } {\bm r}^\prime,
\label{starAccel}
\end{equation} 
which in the cylindrical coordinate system can be given as
\begin{eqnarray}
\Phi_\mathrm{ind}(r_j,\phi_k) & = &  r_{j} \cos(\phi_k) \sum_{j',k'}G {{\bm m}_{j',k'} \over r_j'^2 }\cos(\phi_{k'}) \nonumber \\
& & + \sin(\phi_k) \sum_{j',k'}G {{\bm m}_{j',k'} \over r_j'^2 }\sin(\phi_{k'}),
\label{eq:phi_ind}
\end{eqnarray}
where $m_{j,k}$ and  $r_j$, $\phi_k$, are the mass and cylindrical coordinates of the grid cell $j,k$. To incorporate the effect of disc self-gravity, we calculate the gravitational potential of the disc, $\Phi_{\rm sg}$, by solving for the Poisson integral
\begin{eqnarray} 
  \Phi_\mathrm{sg}(r,\phi) & = & - G \int_{r_{\rm in}}^{r_{\rm out}} r^\prime dr^\prime 
                     \nonumber \\ 
      & &       \times \int_0^{2\pi} 
               \frac{\Sigma(r^\prime,\phi^\prime) d\phi^\prime} 
                    {\sqrt{{r^\prime}^2 + r^2 - 2 r r^\prime 
                       \cos(\phi^\prime - \phi) }}  \, ,
	\label{eq:phi_sg}
\end{eqnarray} 
where $r_{\rm in}$ and $r_{\rm out}$ are the radial positions of the disc inner and outer boundaries. This integral is calculated using an FFT technique which applies the two-dimensional Fourier convolution theorem for polar coordinates logarithmically spaced in the radial direction. The singularity at $r=r'$ and $\phi=\phi'$ is treated by applying the method described in  \citet[][section\ 2.8]{BT87}. In the case of polar coordinates with the log-spaced radial direction, the contribution from the material in the $(j,k)$-th cell to the gravitational potential in the same cell can be calculated analytically (see, Equation 2-206 in \citealp{BT87}) assuming the constant surface density within the $(j,k)$-th cell (which is the usual assumption in the grid-based numerical simulations). This method helps to avoid the problem of singularity without applying gravitational softening. This technique was successfully applied to simulating gravitational instability in embedded protostellar discs (see, e.g., \citealp{VB2010,VB2015}). 

For simplicity, we use the locally isothermal approximation for the gas. This approximation assumes that the thermal heating and cooling processes occur on time-scales that are much faster than the local dynamical period and the disc heating sources do not vary appreciably on time-scales of interest for our modelling. The former is usually fulfilled on radial distances greater than a few au \citep[see fig.~3 in ][]{Vorobyov2014} and the latter is true if stellar luminosity and viscous heating vary weakly with time.

For the disc geometry, we use the flat-disc-surface approximation. In this case the pressure scale-height of the disc has a power-law dependence on the  radius, $H=hR$, where $h$ is the disc aspect ratio assumed to be $h=0.05$. 

To model the formation of a large-scale vortex,  a sharp viscosity transition is introduced at the dead zone outer edge. We assume that the disc has an accretionally inactive region, the dead zone \citep{Gammie1996}, where the value of $\alpha$ is smoothly reduced such that $\alpha_\mathrm{dz}=\alpha\delta_\alpha$. The viscosity reduction is given by 
\begin{equation}
        \label{eq:deltaalpha}
        \delta_\alpha=1-\frac{1}{2}\left(1-\alpha_\mathrm{mod}\right)\left[1-\tanh\left(\frac{R-R_\mathrm{dze}}{\Delta R_\mathrm{dze}}\right)\right],
\end{equation}
where $\alpha_\mathrm{mod}$ is the depth of the turbulent viscosity reduction. We assume that $\alpha=0.01$ in the viscously active disc region. Two scenarios are investigated for which cases $\alpha_\mathrm{mod}=0.01$ and $0.001$, resulting in $\alpha_\mathrm{dz}=10^{-4}$ and $10^{-5}$, respectively.

To quantify the radius of viscosity reduction, $R_\mathrm{dze}=24$\,au is used, where we adopt the results of \citet{MatsumuraPudritz2005}, who found that $R_\mathrm{dze}$ lies between 12\,au and 36\,au. We assume $\Delta R_\mathrm{dze}=1H_\mathrm{dze}$, where $H_\mathrm{dze}=R_\mathrm{dze}h$ is the disc scale
height at the viscosity reduction, which corresponds to $\Delta R_\mathrm{dze}=1.2$\,au. Excitation of the RWI requires a sharp viscosity transition ($\Delta R_\mathrm{dze}\leq2H_\mathrm{dze}$) in the $\alpha$-prescription \citep{Lyraetal2009b,Regalyetal2012}, which is sharper than it is expected to form at the outer dead zone edge \citep{Dzyurkevichetal2013}. We note, however, that  \citet{Lyraetal2015} found that a smooth change in the gas resistivity does not imply an equally smooth transition in the turbulent stress, for which case a large-scale vortex can form although the resistivity transition is  smooth. We note that the total width of the viscosity transition given by Equation\,(\ref{eq:deltaalpha}) is about $2\Delta R_\mathrm{dze}$, which corresponds to 2.4\,au.
 
The unit length is taken to be 1\,au and the unit mass is the stellar mass. Assuming that the unit time is the inverse of the Keplerian frequency, the orbital period becomes 2$\pi$ and the gravitational constant, G, is unity. 

Initially, the gas density has a power-law profile $\Sigma(R)=\Sigma_0R^{-p}$ with $p=1$. We investigate the vortex formation and evolution for three different disc masses: $M_\mathrm{disc}/M_*=0.001,\, 0.005$ and 0.01, in which cases $\Sigma_0=3.18131\times10^{-6},\,1.90986\times10^{-5}$ and $3.18131\times10^{-5}$, respectively. 

The Toomre parameter is defined as $Q=\kappa c_\mathrm{s}/(\pi G\Sigma)$, where $\kappa=(2\Omega_\mathrm{g} r^{-1}d(r^2\Omega_\mathrm{g})/dr)^{1/2}$ is the epicyclic frequency and $\Omega_\mathrm{g}$ is the angular velocity of the pressure supported gas in the gravitational field of the central star \citep{Toomre1964}. Assuming that the gas has  a locally isothermal equation of state, $\Omega_\mathrm{g}=\Omega_{\rm K}[1-p (c_\mathrm{s}/(\Omega_{\rm K} R))^2]^{1/2}$. In a flat-disc-surface approximation $\kappa=\Omega_{\rm K}(1-ph^2)^{1/2}$,  which is $\simeq\Omega_{\rm
K}$ for $h\ll1$, thus Toomre parameter simplifies to $Q=(R^{p-2} h)/(\pi G \Sigma_0)$. The initial values
of the Toomre parameter $Q_\mathrm{init}$ are $\simeq250,\,50$ and $25$ at a distance where the large-scale vortex forms ($R\simeq20$) for the  modelled disc masses of $M_{\rm disk}/M_\ast =0.001, 0.005$ and
0.01, respectively.

The radial and azimuthal velocity components assuming $\alpha$-prescription for the disc viscosity are initially set to
\begin{equation}
        v_r (r,\phi)= -3h^2r^{-1/2}\alpha(1-p),
\end{equation}
\begin{equation}
        v_\phi(r,\phi)=r\sqrt{\Omega_\mathrm{K}(r)^2(1-h^2) (1+p) - \frac{1}{r}\frac{\partial\Phi_\mathrm{sg}(r,\phi)}{\partial r}}.
        \label{eq:v_phi}
\end{equation}
The term $-(1/r)\partial\Phi_\mathrm{sg}(r,\phi)/\partial r$ in Equation\,(\ref{eq:v_phi}) is included to take into account the the radial acceleration due to disc self-gravity.

The spatial extension of the computational domain is $3~\mathrm{au}\leq R\leq50$~au in physical units, which consists of $N_R = 256$ logarithmically  distributed radial (required for the solution of the Poisson integral for disc self-gravity) and $N_\phi=512$ equidistant azimuthal grid cells. With these settings the disc is resolved by $\sim0.2H$ everywhere with approximately square-shaped grid cells. Having compared the results of our simulations with those that adopted a higher numerical resolution (see the Appendix), we can confirm that our simulations are in the numerically convergent regime with the above numerical resolution.

\begin{figure*}
        \centering
        \includegraphics[width=1.8\columnwidth]{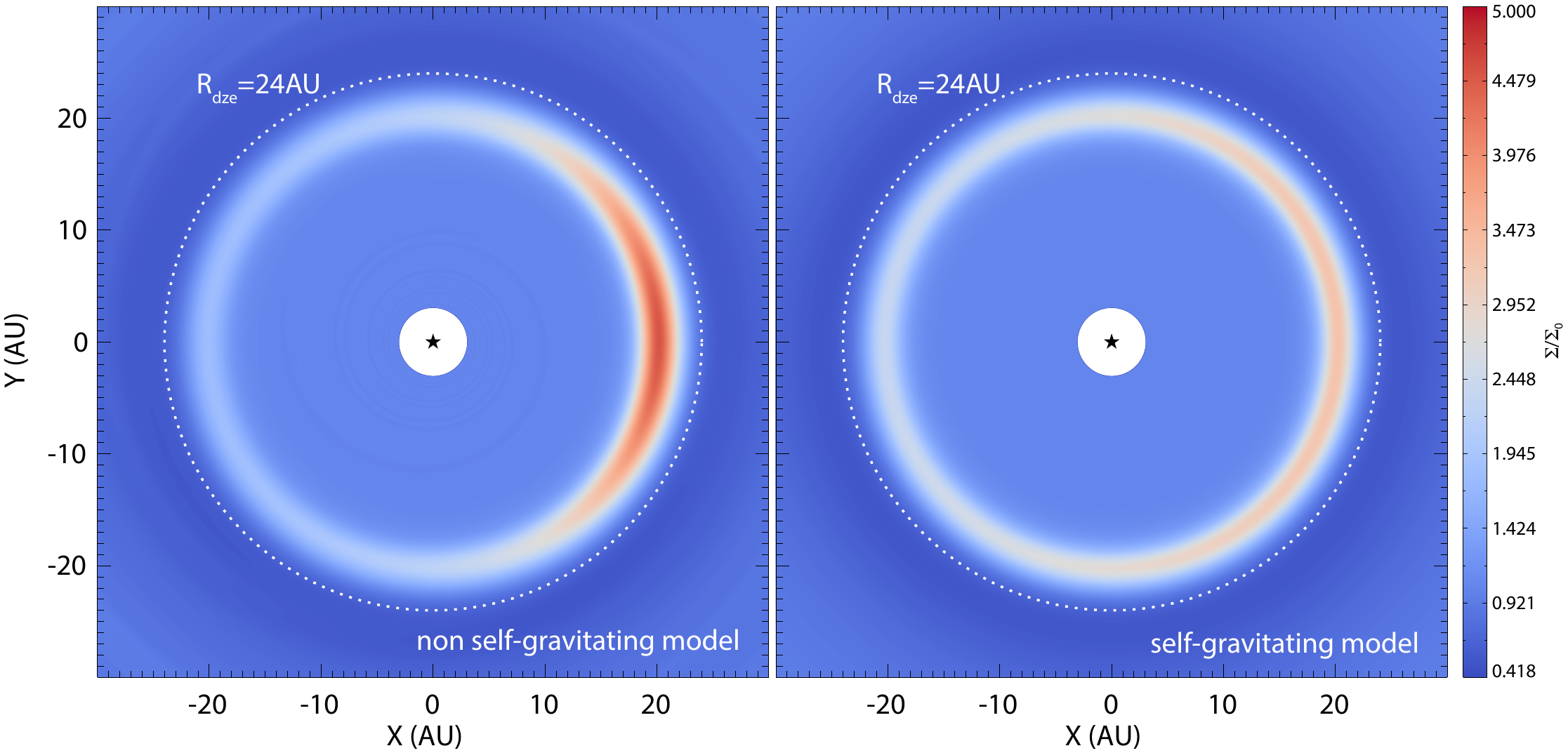}
        \caption{Comparison of the normalized gas density distribution for the non-self-gravitating (left) and self-gravitating (right) models with a disc mass of $M_\mathrm{disc}/M_*=0.005$  and $\alpha_\mathrm{dz}=10^{-4}$ in the dead zone. The snapshots are taken after 500 vortex orbits. The dashed circle shows the dead zone edge.}
        \label{fig1}
\end{figure*}

At the inner and outer boundaries we apply a wave damping boundary condition \citep{deValBorroetal2006}. Note that the disc mass is not conserved strictly with these boundary conditions: the disc mass is increased less than a percent during the simulation. We also note that if we choose the open boundary conditions, this would not affect our results, but the computational time would increase significantly due to waves excited near the inner boundary.  All the simulations cover $1000$ orbits measured at a distance of the vortex centre. 

\section{Results}

\begin{figure*}
        \centering
        \includegraphics[width=2.1\columnwidth]{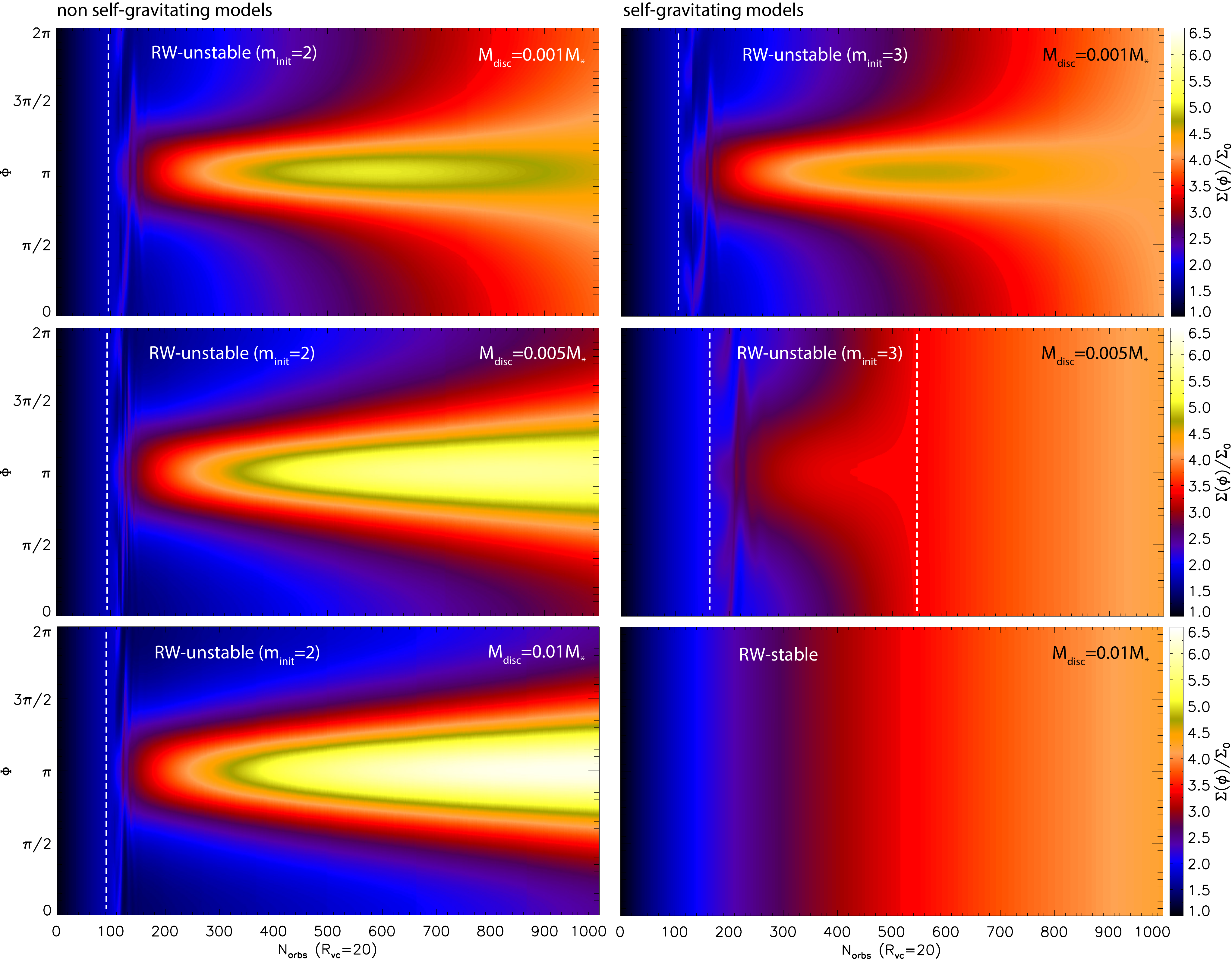}
        \caption{Evolution of the normalized azimuthal density profile measured across the vortex shown as a function of the number of vortex orbits for non-self-gravitating (left) and self-gravitating (right) models assuming $\alpha_\mathrm{dz}=10^{-4}$. Three different disc masses are investigated: $M_\mathrm{disc}/M_*=0.001,\, 0.05$, and $0.01$ from top to bottom. The time of the onset of RWI excitation (indicated by white dashed lines) is independent of the disc mass for the non-self-gravitating models. In contrast, delay is observed proportionally to the disc mass for the self-gravitating models. The fastest growing mode of RWI is 2 for the non-self-gravitating, while 3 for the self-gravitating models. Generally, the non-self-gravitating models have stronger contrast on the profile: $(\Sigma(\Phi)/\Sigma_0)_\mathrm{max}=6.4$ for the non-self-gravitating, while $4.8$ for the self-gravitating models. Note that the disc is Rossby wave (RW) stable throughout the simulation for $M_\mathrm{disc}/M_*=0.01$.}
        \label{fig2}
\end{figure*}

\begin{figure*}
        \centering
        \includegraphics[width=2.1\columnwidth]{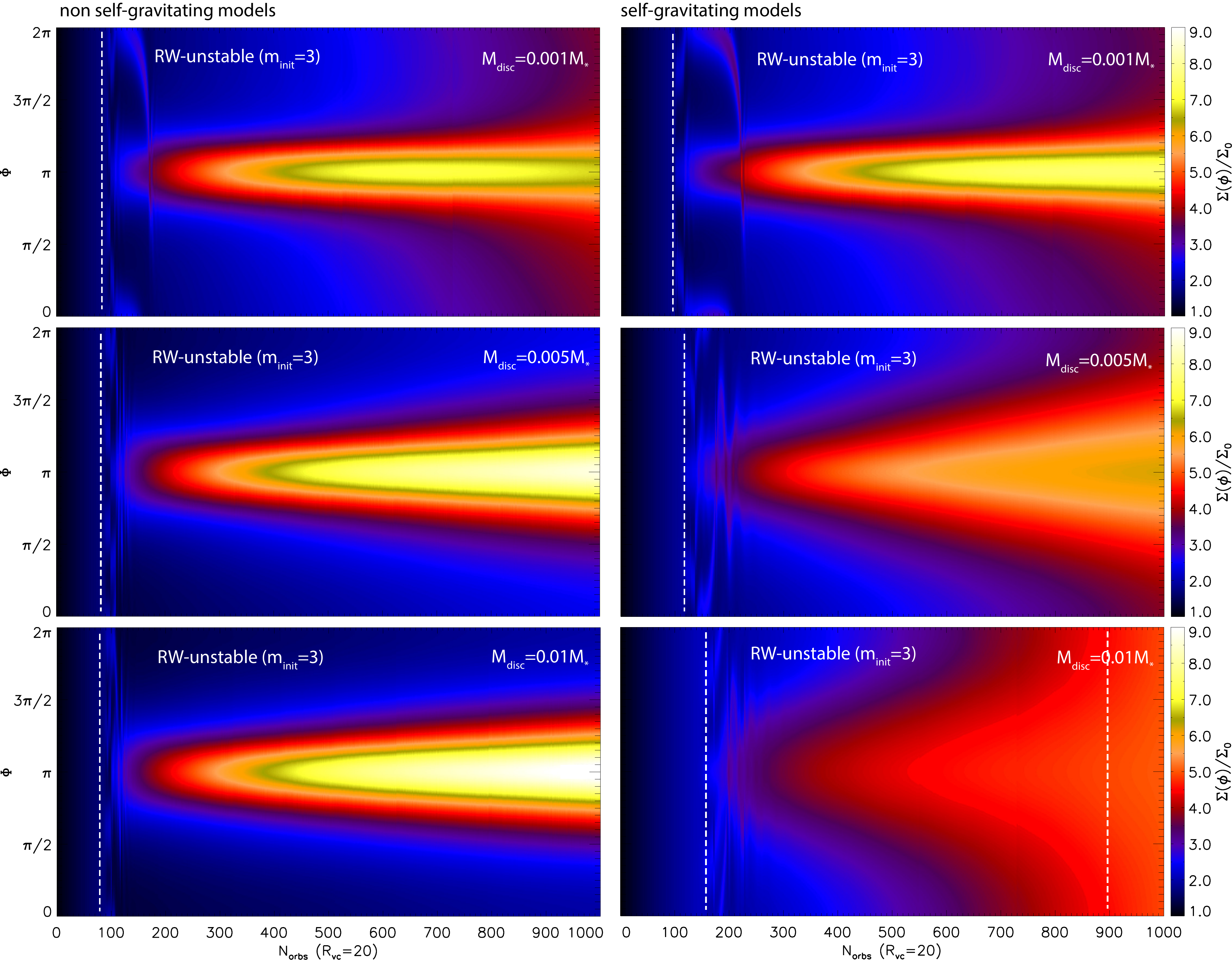}
        \caption{Same as Fig.\,\ref{fig2}, but the dead zone viscosity is $\alpha_\mathrm{dz}=10^{-5}$.  It is appreciable that the vortex dissipation due to viscous evolution is significantly slower than in models shown in Fig.~\ref{fig1}. Contrary to the $\alpha_\mathrm{dz}=10^{-4}$ models, the disc is RW unstable even for $M_\mathrm{disc}/M_*=0.01$. Note that the initially fastest growth mode is 3 independent of whether the disc self-gravity is included or not. Generally, the non-self-gravitating models have stronger contrast on the profile: $(\Sigma(\Phi)/\Sigma_0)_\mathrm{max}=9$ for the non-self-gravitating, while $7.7$ for the self-gravitating models. }
        \label{fig3}
\end{figure*}

\begin{figure*}
        \centering
        \includegraphics[width=2\columnwidth]{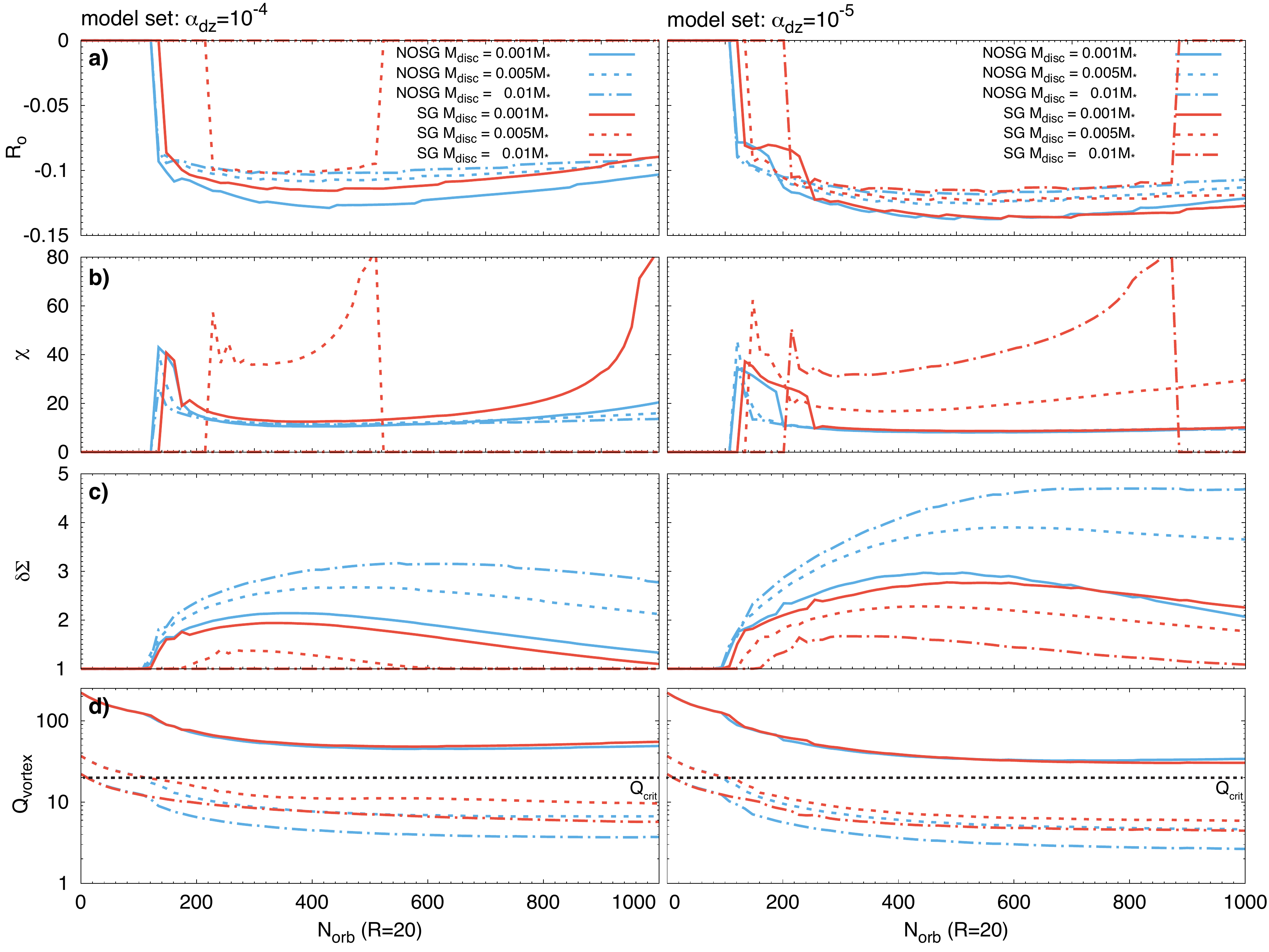}
        \caption{From top to bottom: Rossby number (panel a), $R_\mathrm{o}$, the vortex aspect ratio (panel b), $\chi$,  the azimuthal density contrast (panel c), $\delta\Sigma$, and the Toomre parameter $Q_\mathrm{vortex}$ (panel d), calculated at the vortex centre as a function of time measured by the number of vortex orbits in non-self-gravitating and self-gravitating models. Left hand and right hand panels show $\alpha_\mathrm{dz}=10^{-4}$ and $10^{-5}$ disc models, respectively. If RWI is not excited $R_\mathrm{o}$, $\delta\Sigma$ and $\chi$ are set to 0, 1 and 0, respectively. The critical Toomre $Q_\mathrm{crit}\simeq20$, below which the disc self-gravity is expected to be important in our models is also shown.}
        \label{fig4}
\end{figure*}

\subsection{Vortex formation and evolution}

We observed the excitation of the RWI in all models, except the $\alpha_\mathrm{dz}=10^{-4}$, $M_\mathrm{disc}/M_*=0.01$ model with self-gravity.  Small scale vortices (with a mode number\footnote{The mode number is the largest amplitude component of the Fourier transformation of the azimuthal density profile taken across the vortex, which is equivalent to the number of vortices present in the disc.} $2\leq m\leq3$) develop, which later merge to a single large-scale vortex. Having compared the morphologies of the full-fledged vortices in different models, we found that they are azimuthally more elongated, but have a weaker azimuthal density contrast, if disc self-gravity is included (Fig.~\ref{fig1}).

To explore the long-term evolution of the vortices ($\sim1000$ vortex orbital periods), we first normalized the density distributions with respect to the initial ones. Then, the azimuthal density profiles taken across the vortex centre are calculated. Figs~\ref{fig2} and \ref{fig3} show the evolution of these azimuthal density profiles as a function of the number of orbital periods for two different dead zone viscosities, $\alpha_\mathrm{dz}=10^{-4}$ and $\alpha_\mathrm{dz}=10^{-5}$, respectively. Since the gas continuously accumulates at the viscosity transition, the maxima of the density profiles monotonously increase with time. Concurrently, the vortex widens azimuthally in part due to the viscous evolution of the disc. The effect of disc self-gravity also tends to widen the vortices, which is evident in the right hand panels of Figs~\ref{fig2} and \ref{fig3}. 

For $\alpha_\mathrm{dz}=10^{-4}$ (Fig.~\ref{fig2}), the initial mode number of RWI (i.e. the number of  vortices initially excited) is 2 and 3 for non-self-gravitating and self-gravitating cases, respectively. We emphasise that no RWI excitation is observed for the most massive ($M_\mathrm{disc}/M_*=0.01$) model if disc self-gravity is included. For $\alpha_\mathrm{dz}=10^{-5}$ (Fig.~\ref{fig3}), the RWI is excited in all models with the $m=3$ mode independent of the disc mass and the inclusion of disc self-gravity.

For the non-self-gravitating models, the RWI is excited after $\sim100$ orbital periods independent of the disc mass. However, for the self-gravitating models, the RWI excitation is delayed proportionally to the disc mass: RWI is excited after $\sim100,\,150$ and $\sim180$ orbital periods for the $M_{\rm d}=0.001,\,0.005$ and $0.01\,M_*$ discs, respectively.  

We found that the smaller the dead zone viscosity, the larger the vortex azimuthal contrast. Moreover, this contrast is generally stronger for the non-self-gravitating models. The full-fledged vortex is also found to decay faster in the self-gravitating models (see the density contrast across the azimuth in Figs~\ref{fig2} and \ref{fig3}). The vortex is sustained till the end of the simulation for the non-self-gravitating models, while it is completely dissolved by $\sim600$ orbits for the $M_\mathrm{disc}/M_*=0.005$, $\alpha_\mathrm{dz}=10^{-4}$ self-gravitating models.

\subsection{Vortex strength and shape}

To compare the vortex strength and shape, we calculate the time evolution of three vortex properties in the non-self-gravitating and the self-gravitating models: the Rossby number, $R_\mathrm{o}$, the vortex aspect ratio, $\chi$, both characterizing the vortex strength, and the azimuthal density contrast, $\delta\Sigma$, affecting the vortex dust accumulation efficiency. The Rossby number being the $z$ component of the vorticity in the local frame of the vortex divided by the global vorticity of the Keplerian disc calculated as
\begin{equation}
        R_\mathrm{o}(R,\phi)=\frac{\nabla\times({\bm v}(R,\phi)-R\Omega_\mathrm{K}{(R)})}{2\Omega_\mathrm{g}},
        \label{eq:Rossby-number}
\end{equation}
where $\Omega_\mathrm{g}$ is the angular velocity of the gas at the vortex centre defined  at the vortensity minimum.  Assuming that the flow pattern inside the vortex is an ellipse (see, e.g.  \citealp{Kida1981,Chavanis2000}), $\chi$ is defined by the ratio of the azimuthal and radial axes of the ellipse fitted to 2D vortex density field. $\delta\Sigma$ is measured as the ratio of the maximum and the minimum values of the azimuthal density profile taken across the vortex. 

Fig.~\ref{fig4} shows the Rossby number (panel a), vortex aspect aspect ratio (panel b), vortex azimuthal contrast (panel c), and Toomre parameter, $Q_\mathrm{vortex}$, calculated at the vortex centre or at the density maximum in models where no RWI develops (panel d) against time. The full-fledged $m=1$ large-scale vortex is formed with $R_\mathrm{o}\simeq-0.1$ and $\chi\simeq40$  in models where RWI is excited. Note that the $m=1$ mode vortex does not develop in the self-gravitating, $\alpha=10^{-4}$ model for $M_\mathrm{disc}/M_*=0.01$, thus $R_\mathrm{o}=0$ and $\chi=0$ throughout the simulation. In the corresponding $\alpha_\mathrm{dz}=10^{-5}$ model RWI is excited, thus $R_\mathrm{o}$ and $\chi$ are non-zero, but vanishes before the end of the simulation at about 900 orbits. Moreover, in the $M_\mathrm{disc}/M_*=0.005$ self-gravitating, $\alpha=10^{-4}$ model, the vortex also dissolves before the end of the simulation, thus $\chi$ becomes zero after $\sim500$ orbits.

After a full-fledged $m=1$ vortex developed, $R_\mathrm{o}$ and $\chi$ starts to decline as the vortex strengthens. Later, $R_\mathrm{o}$ and $\chi$ reaches their minima ($R_\mathrm{o,min}\simeq-0.13$, $\chi_\mathrm{min}\simeq10$ and $R_\mathrm{o,min}\simeq-0.14$, $\chi_\mathrm{min}\simeq8$ for $\alpha_\mathrm{dz}=10^{-4}$ and $10^{-5}$, respectively). The magnitude of $R_\mathrm{o,min}$ is smaller for larger disc mass independent of whether the self-gravity is included or not. This means that the vortensity is weaker inside the vortex at the vortex's strongest state for the more massive discs.  Although a weaker vortex has a larger vortex aspect ratio, a significant change in $\chi_\mathrm{min}$ can only be observed in self-gravitating models.  This might be due to that the Kida description is not valid or the vortex is very weak, in these cases the $R_\mathrm{o,min}-\chi$ function is flattened, but see our analysis on the vortex models in Section~\ref{sect:vortex_models}.

Subsequently, the vortex weakens with time in all models, however, the evolutions of $R_\mathrm{o}$ and $\chi$ are very slow and only weakly dependent on the disc mass in non-self-gravitating discs, especially in $\alpha_{dz}=10^{-5}$ non-self-gravitating models. In contrast, the evolutions of $R_\mathrm{o}$ and $\chi$ are strongly dependent on the disc mass for self-gravitating models: the larger the disc mass, the faster the decline in $R_\mathrm{o}$ and $\chi$, i.e. the vortex strength decreases faster in a more massive disc.

Surprisingly, the azimuthal density contrast, $\delta\Sigma$, shows a very strong dependence on the disc mass in non-self-gravitating as well as in self-gravitating models. Note that $\delta\Sigma$ is unity if no vortex is present, i.e. for $M_\mathrm{disc}/M_*=0.01$ and $0.005\,M_*$ self-gravitating models.  $\delta\Sigma$ starts to increase after the full-fledged $m=1$ vortex formed and reaches a certain maximum as it strengthens. $\delta\Sigma_\mathrm{max}$ is proportional to the disc mass for the non-self-gravitating models, while an opposite trend can be observed for the self-gravitating models. More specifically,  $\delta\Sigma_\mathrm{max}\lesssim3$ for $\alpha_\mathrm{dz}=10^{-4}$ and $\delta\Sigma_\mathrm{max}\lesssim5$ for $\alpha_\mathrm{dz}=10^{-5}$ in the non-self-gravitating models, while $\delta\Sigma_\mathrm{max}\lesssim2$ for $\alpha_\mathrm{dz}=10^{-4}$ and $\delta\Sigma_\mathrm{max}\lesssim3$ for $\alpha_\mathrm{dz}=10^{-5}$ in the self-gravitating models. Note that the maximum contrast develops later for a more massive disc in non-self-gravitating, while sooner in self-gravitating models.,Subsequently, $\delta\Sigma$ starts to decline with a rate being larger(smaller) for a lower mass disc for the non-self-gravitating(self-gravitating) models. 

A general tendency in the evolution of $Q_\mathrm{vortex}$ can be seen in panel~(d) of Fig.\,\ref{fig4} -- $Q_\mathrm{vortex}$ decreases with time as gas continuously accumulates at the viscosity transition, and later $Q_\mathrm{vortex}$ reaches a constant value as a quasi-steady state forms by the end of the simulations. Disc self-gravity has no effect on the evolution of $Q_\mathrm{vortex}$ for the least massive ($M_\mathrm{disc}/M_*=0.001$) models. However,  $Q_\mathrm{vortex}$ decreases with a slower rate in $M_\mathrm{disc}/M_*=0.005$ and $0.01$ self-gravitating models after the full-fledged, $m=1$-mode large-scale vortex is formed. In agreement with \citet{LovelaceHohlfeld2013}, a notable difference in the time
evolution of $Q_\mathrm{vortex}$ between non-self-gravitating and self-gravitating discs occurs 
at $Q_\mathrm{vortex}<Q_\mathrm{crit}\simeq20$.

\section{Discussion}
 
 \begin{figure*}
        \centering
        \includegraphics[width=2\columnwidth]{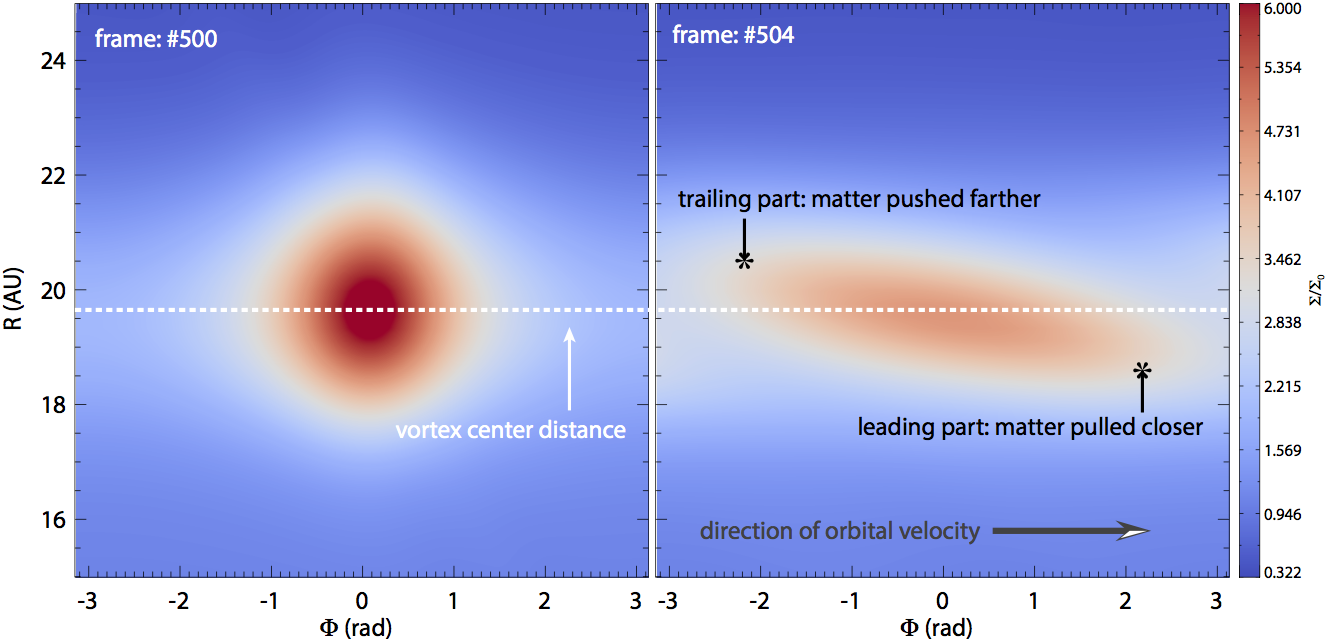}
        \caption{Close-up of the large-scale vortex shown in cylindrical coordinate system prior to (left) and after 4 vortex orbits the self-gravity is turned on (right). $M_\mathrm{disc}/M_*=0.01$ and $\alpha_\mathrm{dz}=10^{-5}$ in this model. It is appreciable on the right hand panel that the vortex become stretched and tilted such that the leading and trailing parts of the vortex orbit at smaller and larger stellar distances, respectively.}
        \label{fig5}
\end{figure*}

 \subsection{Vortex morphology and evolution}
 
One can see in Fig.\,\ref{fig2} that the excitation of RWI excitation begins with $m=2$ for the non-self-gravitating and $m=3$ mode number for self-gravitating $\alpha_\mathrm{dz}=10^{-4}$ models. For $\alpha_\mathrm{dz}=10^{-5}$ models, however, $m=3$ mode is excited initially, independent of that whether disc self-gravity is included or not (Fig.\,\ref{fig3}). Moreover, the formation of the full-fledged $m=1$ vortex is prolonged for later times if the disc self-gravity is taken into account. According to \citet{LinPapaloizou2011}, the fastest growing mode of the RWI is shifted to large mode numbers by disc self-gravity, which we confirm for dead zone edge vortices. \citet{LinPapaloizou2011} and \citet{Lin2012} also showed that disc self-gravity prolongs or even suppresses vortex development at the vicinity of gap edges opened by a giant planet for a disc with the Toomre parameter $\lesssim10$. A similar phenomenon occurs for the dead zone edge vortices if the Toomre parameter at the vortex centre ($Q_\mathrm{vortex}$) drops below a critical of $Q_\mathrm{crit}=1/h\simeq20$ (assuming h=0.05) reported by \citet{LovelaceHohlfeld2013}.
 
The vortex weakening is slower, therefore its lifetime is longer for a more massive disc in the non-self-gravitating models, while the opposite trend is observed in the self-gravitating models (see, e.g., Figures\,\ref{fig2} and \ref{fig3}). On one hand, this is caused by the fact that the full-fledged vortex is stronger(weaker) for a more massive disc in the non-self-gravitating(self-gravitating) models. On the other hand, the rate of vortex decay is also found to be correlated with the disc mass in the same way.  As a consequence, vortices developed in a more massive discs tend to live for shorter time due to the disc self-gravity. 

We have also seen that $\delta\Sigma_\mathrm{max}$ is proportional to the disc mass in the non-self-gravitating models (panel c of Figure\,\ref{fig4}). This phenomenon can be explained by the effect of the indirect potential, $\Phi_\mathrm{ind}(r,\phi)$, generated by the azimuthal density asymmetry representing the vortex itself.  \citet{MittalChiang2015} showed that the displacement of the barycentre of the disc-star system from the star results in the formation of a lopsided disc for $M_\mathrm{disc}/M_*\gtrsim0.01$. \citet{ZhuBaruteau2016} found that a large-scale vortex generated by an artificial pressure bump in a massive disc ($M_\mathrm{disc}/M_*\gtrsim0.01$) becomes spatially more concentrated if the indirect potential is taken into account (see their fig.\,3). Recently, we also showed that a vortex formed at a sharp viscosity transition is more concentrated and has a longer lifetime if the displacement of the barycentre caused by the vortex itself, i.e. the indirect potential is correctly taken into account \citep{RegalyVorobyov2017}. Since the strength of the indirect potential being proportional to the disc mass (see Equation\,\ref{eq:phi_ind}), $\delta\Sigma_\mathrm{max}$ becomes larger in a more massive disc due to a larger displacement of the barycentre of the system. 

In contrast, if self-gravity is taken into account, $\delta\Sigma_\mathrm{max}$ is inversely proportional to the disc mass independent of the value of the viscosity applied (see lower panel on Figure\,\ref{fig4}). Along with this $\chi_\mathrm{min}$ increases with the disc mass for self-gravitating discs (see middle panel on Figure~\,\ref{fig4}). This can be explained by the fact that the full-fledged vortices are stretched azimuthally in self-gravitating models compared to the cases in the non-self-gravitating models (see Figure\,\ref{fig1}). 

As a result of disc self-gravity, the decline of $Q_\mathrm{vortex}$ (see panel d of Figure\,\ref{fig4}) is somewhat modest compared to that of non-self-gravitating disc, resulting in $Q_\mathrm{vortex}^{SG}>Q_\mathrm{vortex}^{NOSG}$ inside the full-fledged $m=1$ vortex. Since $Q\sim1/\Sigma$,  the density enhancement in the vortex centre is weaker for self-gravitating discs. This phenomenon is counterintuitive, however, it can be explained by the vortex stretching effect of disc self-gravity discussed in the next section.

 \subsection{Vortex stretching by gravitational torque}

To understand the vortex stretching phenomenon, we performed a controlled numerical experiment by taking a non-self-gravitating model and turning on disc self-gravity after the vortex becomes fully developed.  For this experiment, we selected the most massive $M_\mathrm{disc}/M_*=0.01$ model with $\alpha=10^{-5}$. The left hand panel in Fig.\,\ref{fig5} shows the full-fledged vortex in the non-self-gravitating model at $500$ vortex orbits. The right hand panel in  Fig.\,\ref{fig5} shows a strongly stretched and tilted vortex, which is formed $\sim4$ orbital periods after self-gravity is turned on. In the latter case, the radial distance of the vortex centre is unchanged, but the trailing and leading parts of the vortex wings (with respect to the vortex centre and in the direction of the gas orbital motion) are pulled back and pushed forward, respectively. Note that the vortex stretches with the same rate for the $\alpha=10^{-4}$ model.

\begin{figure*}
        \centering
        \includegraphics[width=2\columnwidth]{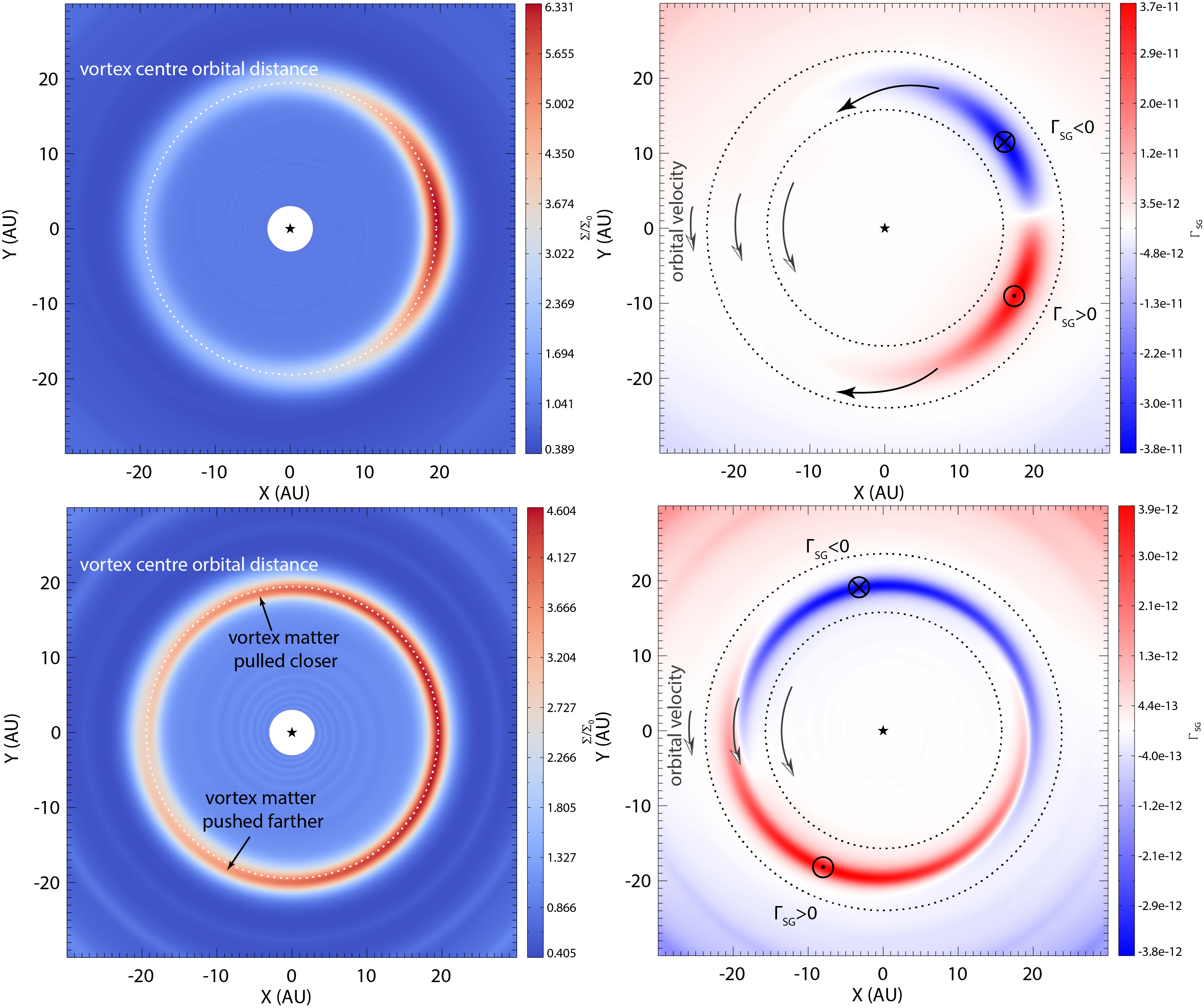}
        \caption{Stretching of the large-scale vortex in the self-gravitating disc is started immediately after the self-gravity is turned on. Upper row shows the disc prior to turning on the self-gravity, while the lower row shows the disc after  $\sim$4 vortex orbits. The disc's gravitational torque distribution, $\Gamma_\mathrm{SG}(r,\phi)$, calculated by Equation~(\ref{eq:gravTorque}) is shown in the right hand panels. The gas parcels lose(gain) angular momentum causing orbital decay(growth) at the vortex wings where the gravitational torque is negative(positive). Since the orbital velocity is higher(lower) at smaller(larger) orbital distances, the leading and the trailing part recede from each other. As a result, the vortex is stretched and eventually decayed in this particular model.}
        \label{fig6}
\end{figure*}

The vortex stretching and tilting can be explained by the effect of the gravitational torque exerted by the vortex itself. The gravitational torque exerted on a given cell with coordinates ($r, \phi$) in the cylindrical coordinate system can be calculated as
\begin{equation}
 \Gamma_\mathrm{sg}(r,\phi)={\bm r}\times\Sigma(r,\phi)A(r,\phi)\nabla\Phi_\mathrm{sg}(r,\phi),
\label{eq:Gamma_sg}
\end{equation}
where $A(r,\phi)$ is the surface area of a cell with coordinates ($r,\phi$).
The only non-zero component of ${\bf \Gamma_\mathrm{sg}}(r,\phi)$ is the azimuthal component\footnote{
Here, we neglected the possible (small) deviation of the centre of mass from the coordinate centre.}, which is
\begin{equation}
\Gamma_\mathrm{sg,\phi}(r,\phi)=\Sigma(r,\phi)A(r,\phi)\frac{\partial\Phi_\mathrm{sg}(r,\phi)}{\partial \phi}.
\label{eq:gravTorque}
\end{equation}
Fig.\,\ref{fig6} shows the spatial distributions of the gas surface density and gravitational torque calculated according to Equation (\ref{eq:gravTorque}) at the moment when disc self-gravity is turned on and after the vortex has significantly stretched. Evidently, the leading part of the vortex is characterized by the negative gravitational torque, which removes angular momentum from the gas in the leading wing, forcing its gas parcels to move closer to the star.  In a Keplerian disc, it also means that these gas parcels will speed up, effectively stretching the vortex in the forward direction. Conversely, the gravitational torque at the trailing part of the vortex is positive, thus depositing angular moment to the trailing wing and forcing its gas parcels to increase their orbital distance. Since the gas orbital velocity is proportional to $r^{-0.5}$, the leading and trailing wings of the vortex are receding from each other, i.e. the vortex stretches and gets tilted. Due to this vortex stretching, the azimuthal asymmetry decreases causing the magnitude of the disc's gravitational torque to fall by about an order of magnitude during  $\sim 4$ orbits. 

Since the magnitude of the gravitational torque is proportional to the disc mass, see Equation~(\ref{eq:gravTorque}), the full-fledged vortex decays faster for more massive discs. Therefore, the vortex's gravitational torque tends to weaken the full-fledged vortex and also fasten the vortex decay. Consequently, the vortex's gravitational torque tends to decrease the lifetime of the dead-zone-edge vortex  proportionally to the disc mass.

The vortex stretching due to disc self-gravity is counterintuitive. One may think that self-gravity should help gravitational contraction of the vortex, rather than its dissipation. In this phenomenon, the Keplerian shear is the key factor that acts to stretch the vortex when its leading/trailing ends are decelerated/accelerated due to the gravitational torques exerted by the vortex itself. For the $Q_\mathrm{init}\gg1.0$ and $Q_{\rm vortex} \gtrsim4.0$ cases we investigated, the Keplerian shear (and the restoring pressure force) dominates the effect of gravitational contraction due to self-gravity. That is why the effect of self-gravitational contraction  is overwhelmed by stretching. In the opposite case of massive discs with $Q_\mathrm{init}\simeq1$, the situation may be more complicated  because self-gravity is supposed to dominate over the Keplerian shear, but the gravitational torques will also be stronger, leading to stronger deceleration/acceleration of the leading/trailing ends of the vortex. This can act to increase the radial and azimuthal stretching of the vortex, thus counterbalancing to some extent the contracting effect of self-gravity. This case requires a focused investigation. 

 \subsection{Theoretical vortex models}
\label{sect:vortex_models}

For a steady state incompressible elliptic (with an aspect ratio of $\chi$)  vortex having uniform vorticity, the Rossby number can be approximated as
\begin{equation}
       R_\mathrm{o}^\mathrm{Kida}=-\frac{3}{4}\frac{\chi^2+1}{\chi(\chi-1)}+\frac{3}{4}
       \label{eq:Kida}
\end{equation}
according to \citet{Kida1981}. Assuming a nearly linear spatial dependance of the velocity, i.e. the vorticity field, \citet{Goodmanetal1987} proposed the so-called GNG model for approximating the Rossby number, which reads
\begin{equation}
        R_o^\mathrm{GNG}=-\frac{\sqrt{3}}{2}\frac{\chi^2+1}{\chi\sqrt{\chi^2-1}}+\frac{3}{4}.
        \label{eq:GNG}
\end{equation}
Recently, \citet{SurvilleBarge2015} proposed a more sophisticated vortex model which takes into account the transition between the inner part of the vortex and the background flow, radially stratified density and the temperature backgrounds, and compressional effects of strong vortices. Their Gaussian model gives the following expression for the Rossby number
\begin{equation}
        R_o^\mathrm{Gaussian}=\frac{1}{2}\frac{\chi^2+1}{\chi^2-1}(\frac{3}{2}-\sqrt{3})-\frac{3}{2}.
        \label{eq:Gauss}
\end{equation}

\begin{figure}
        \centering
        \includegraphics[width=\columnwidth]{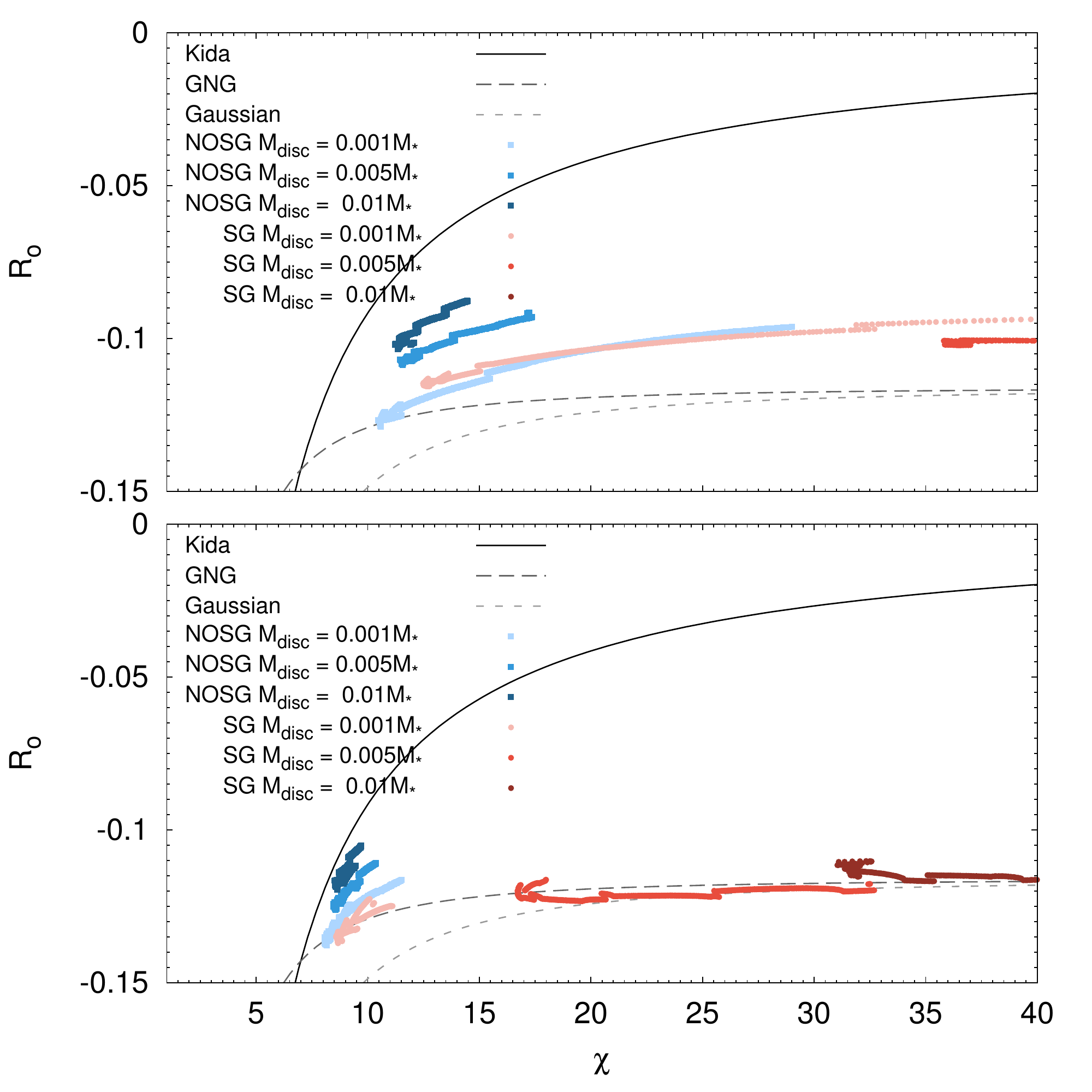}
        \caption{Rossby number and vortex aspect ratio measured in the simulations assuming $\alpha_\mathrm{dz}=10^{-4}$ (upper panel) and $\alpha_\mathrm{dz}=10^{-5}$ (lower panel). The analytic approximation of $R_\mathrm{o}(\chi)$ functions for Kida, GNG and Gaussian models are also shown with solid, long-dashed and dashed lines, respectively.}
        \label{fig7}
\end{figure}

Fig.~\ref{fig7} shows the Rossby numbers determined according to Equation~(\ref{eq:Rossby-number}) as a function of the vortex aspect ratios determined by fitting ellipses to density distribution close to the vortex centre for all models. The above discussed three theoretical models are also shown. On one hand, the vortices are weaker, i.e. the magnitudes of $R_\mathrm{o}$ are lower than those predicted by GNG and Gaussian models for the non-self-gravitating discs. Note that Equations~(\ref{eq:GNG}) and (\ref{eq:Gauss}) give $R_o=(1/4)(3-2\sqrt{3})\lesssim-0.116$, when $\chi\rightarrow\infty$. The Kida model, however, fits the measurements only for the most massive non-self-gravitating $\alpha_\mathrm{dz}=10^{-5}$ models as the vortices are stronger in $\alpha_\mathrm{dz}=10^{-4}$ models then that predicted by the Kida approximation. On the other hand, the self-gravitating models can be described well by the GNG model for $\alpha=10^{-5}$, while it gives somewhat stronger vortices than what is observed in the simulations for $\alpha_\mathrm{dz}=10^{-4}$. 

Generally, if disc self-gravity is neglected, Kida model is applicable for vortices formed at viscosity transition as long as the disc mass is relatively large ($M_\mathrm{disc}/M_*\gtrsim0.01$) and for a nearly inviscid ($\alpha_\mathrm{dz}\lesssim10^{-5}$) disc dead zone. However, for self-gravitating discs the measured vortex aspect ratios and the Rossby numbers are inconsistent with both Kida and Gaussian models, while the GNG model seems to be appropriate as long as the gas viscosity is low ($\alpha_\mathrm{dz}\lesssim10^{-5}$). 

Note that the above discussed theoretical vortex models implicitly assume that the disc gas is inviscid, which can be responsible for their inadequacy in describing vortex evolution in viscous (e.g, for $\alpha_\mathrm{dz}\gtrsim10^{-4}$) discs.  For this reason, the development of a  model for vortices formed at viscosity transitions which takes into account the gas viscosity would be desirable, however, it is beyond the scope of the current study.

\section{Conclusions}

In this paper, we explore the effect of disc self-gravity on the long-term evolution of large-scale vortices developed at a sharp viscosity transition placed at $R_\mathrm{dze}=24$\,au by means of two-dimensional numerical hydrodynamic simulations. The disc thermodynamics is modelled in the locally isothermal approximation, and the $\alpha$-prescription is used for the gas viscosity. We considered $M_\mathrm{disc}/M_*=0.001,0.005$, and $0.01$ disc models with the initial values of the Toomre parameter $Q_\mathrm{init}\simeq$250, 50, and 25 at $R\simeq20$\,au distance where the full-fledged vortex develops.  According to \citet{LovelaceHohlfeld2013}, disc self-gravity is expected to be important only for the models with $Q\leq Q_\mathrm{crit}=1/h$, which for a canonical disc aspect ratio of $h=0.05$ corresponds to $Q_\mathrm{crit}=20$ at a distance where the RWI is excited in our models. We demonstrated that disc self-gravity must be taken into account when  modelling the formation of a large-scale vortex at viscosity transitions in discs where $Q_\mathrm{init}\lesssim50$ because the Toomre parameter at the vortex centre drops below the critical value during the vortex formation. Our main findings are the following.

\noindent
1) We observed a delayed vortex formation and a weaker vortex at a sharp viscosity transition, if disc self-gravity is taken into account similarly to what was found for the gap-edge vortices by \citet{LinPapaloizou2011} and \citet{Lin2012}. 

\noindent
2) Concerning the vortex morphology, we found that the full-fledged vortex becomes azimuthally elongated in self-gravitating discs (Fig.\,\ref{fig1}). The aspect ratio of the full-fledged vortex $\chi$ is proportional to the disc mass in self-gravitating discs and its value at the strongest stage of the vortex lies in the $10\lesssim\chi_\mathrm{min}\lesssim40$ limits. In contrast,  $\chi_\mathrm{min}\simeq10$ is independent of the disc mass in non-self-gravitating discs. The azimuthal density contrast $\delta\Sigma_\mathrm{max}$ across 
the vortex is also sensitive to the disc mass and $\sim50$ percent lower if self-gravity is included. In particular, $\delta\Sigma_\mathrm{max}\lesssim 3$ and 2 for $\alpha_{dz}=10^{-4}$ and $\delta\Sigma_\mathrm{max}\lesssim5$ and 3 for $\alpha_{dz}=10^{-5}$ in the non-self-gravitating and self-gravitating models, respectively.

\noindent
3) The full-fledged vortex is subject to a decay not only due to disc viscosity but also due to disc self-gravity. Moreover, disc self-gravity accelerates the vortex decay as compared to the non-self-gravitating case. The rate of the vortex decay is proportional to the disc mass and becomes significant for $M_{\rm disc}\gtrsim0.005\,M_*$, where $Q_\mathrm{init}\lesssim50$ at the vortex radial distance in the unperturbed disc phase.

\noindent
4) The accelerated vortex decay can be explained by azimuthal stretching of the vortex caused by the vortex's non-vanishing gravitational torque and the Keplerian shear of the disc. Since the magnitude of the vortex gravitational torque is proportional to the disc mass, the vortex lifetime decreases with increasing disc mass. If disc self-gravity is neglected, an opposite correlation is observed between the vortex lifetime and the disc mass, which can be explained by the displacement of the barycentre of the star-disc system caused by the vortex itself \citep{MittalChiang2015,ZhuBaruteau2016,RegalyVorobyov2017}. 

\noindent
5) Finally, we found that vortices developed at sharp viscosity transitions of self-gravitating discs can be well described by the GNG model \citep{Goodmanetal1987} as long as the disc viscosity is low, i.e for $\alpha_\mathrm{dz}\lesssim10^{-5}$.

\subsection{Caveats}

Here we mention some caveats of our models. First, we have adopted the two-dimensional, thin-disc approximation. It is known that in three dimensional models the non-self-gravitating Kida vortices with $\chi\lesssim4$ are subject to the elliptic instability  \citep{LesurPapaloizou2009}, which can destroy the vortex. Although vortices formed by the RWI are found to be similar in three- and two-dimensional simulations with neglected disc self-gravity \citep{Meheutetal2010,Meheutetal2012c}, the vortex aspect ratio can be different if disc self-gravity is included. We emphasize that disc self-gravity tends to increase $\chi$ in two-dimensional simulations, i.e., disc self-gravity might stabilize the three-dimensional vortex against the elliptic instability.  Also note that no gravitational softening is applied for solving Equation~(\ref{eq:phi_sg}),  thus the effect of self-gravity may be over-estimated compared to an equivalent three-dimensional disc.

We assume a locally isothermal disc. However, the equation of state of the gas can be different, which can alter the mass accumulated inside the vortex. As a result, the disc gravitational torque may be altered, which can influence the vortex stretching and the vortex lifetime. Thus, it would be desirable to model the thermodynamics of  self-gravitating discs assuming a more elaborated equation of state for the gas.

We neglect the presence of dust in the disc, which can also influences the evolution of the vortex. It is known that dust accumulation might destroy the vortex via the effect of the dust feedback if the dust-to-gas mass ratio is well above unity, as was observed in non-self-gravitating simulations \citep{Fuetal2014b,Mirandaetal2017}. Since the effect of self-gravity tends to weaken the vortex, the dust-to-gas mass ratio may not grow to high values. Therefore, it is worth investigating by simultaneously taking into account the dust feedback and the effect of disc self-gravity.

We implicitly assume that the position of the dead zone edge is stationary. In reality, the dead zone edge distance might change in time. \citet{Matsumuraetal2008} found that  the dead zone outer edge moves inward with time as the density jump at dead zone develops. Moreover,  in long-term simulation the disc mass loss via photoevaporation or disc wind might be important.

\subsection{Outlook}

Our proposed mechanism works globally in both Toomre unstable and stable discs, because it is not related to gravitational instability, but rather to gravitational torques which naturally occur in any non-axisymmetric discs, including those with large-scale vortices.  Based on our results, we conclude that the formation of long-lasting vortices (e.g., more than a thousand orbital periods) requires a relatively small disc-to-star mass ratios being less than $0.5$ percent and low disc viscosity in the dead zone, $\alpha\lesssim10^{-4}$. However, to find the possible disc configurations that favour long-lasting vortices requires a more detailed parameter study.

Large-scale vortices formed at the edges of a gap opened by an embedded planet \citep{Lietal2005} or at the edges of the disc's accretionally inactive dead zone \citep{Lovelaceetal1999, Lietal2000} can explain horseshoe-shaped brightness asymmetries observed for several transitional discs \citep{Regalyetal2012}. The two formation scenarios, however, results in different vortex morphology: while gap edge vortices are azimuthally concentrated, dead zone edge vortices are azimuthally more extended. This phenomenon can be used to infer the vortex formation scenario (Reg\'aly, in preparation). Since disc self-gravity tends to increase the vortex azimuthal extension and decrease the azimuthal contrast affecting its dust collection efficiency, it may have a significant effect on the observed brightness asymmetries. Thus, it is worth investigating the long-term evolution of gap edge vortices and also the dust collection efficiency taking into account the disc's self-gravity.

According to a previously proposed scenario, large-scale vortices can become gravitationally unstable and collapse to massive planets, although the disc itself may be gravitationally stable (see, e.g., \citealp{AdamsWatkins1995,LinPapaloizou2011}). We conclude that this can not be a plausible pathway to planet formation unless the disc is very massive (i.e. gravitationally unstable), because of the vortex stretching caused by disc self-gravity, which tends to decrease the density enhancement at the vortex centre.  Another promising planet formation scenario could be the vortex-aided planet formation, wherein vortices behave as planetary cradles by accumulating significant amount of dust. Our results show that this mechanism would favour low-mass and low-viscosity protoplanetary discs, or late phases, when the disc already lost significant amount of its mass.

\section*{Acknowledgements}

This project was supported by the Hungarian OTKA Grant No. 119993.
Zs. Regaly acknowledges support from the MTA CSFK Lend\"ulet disc Research Group.
E. Vorobyov acknowledges support from the Russian Science Foundation grant 17-12-01168.
We gratefully acknowledge the support of NVIDIA Corporation with the donation of the Tesla 2075 and K40 GPUs. We also acknowledge NIIF for awarding us access to computational resource based in Hungary at Debrecen.  We thank the anonymous referee for insightful comments and suggestions to improve the quality of  the paper.

\bibliographystyle{mn2e}

\bibliography{mn-jour,regaly}

\appendix

\section{Numerical convergency}
\label{sec:apx-1}

\begin{figure}
        \centering
        \includegraphics[width=\columnwidth]{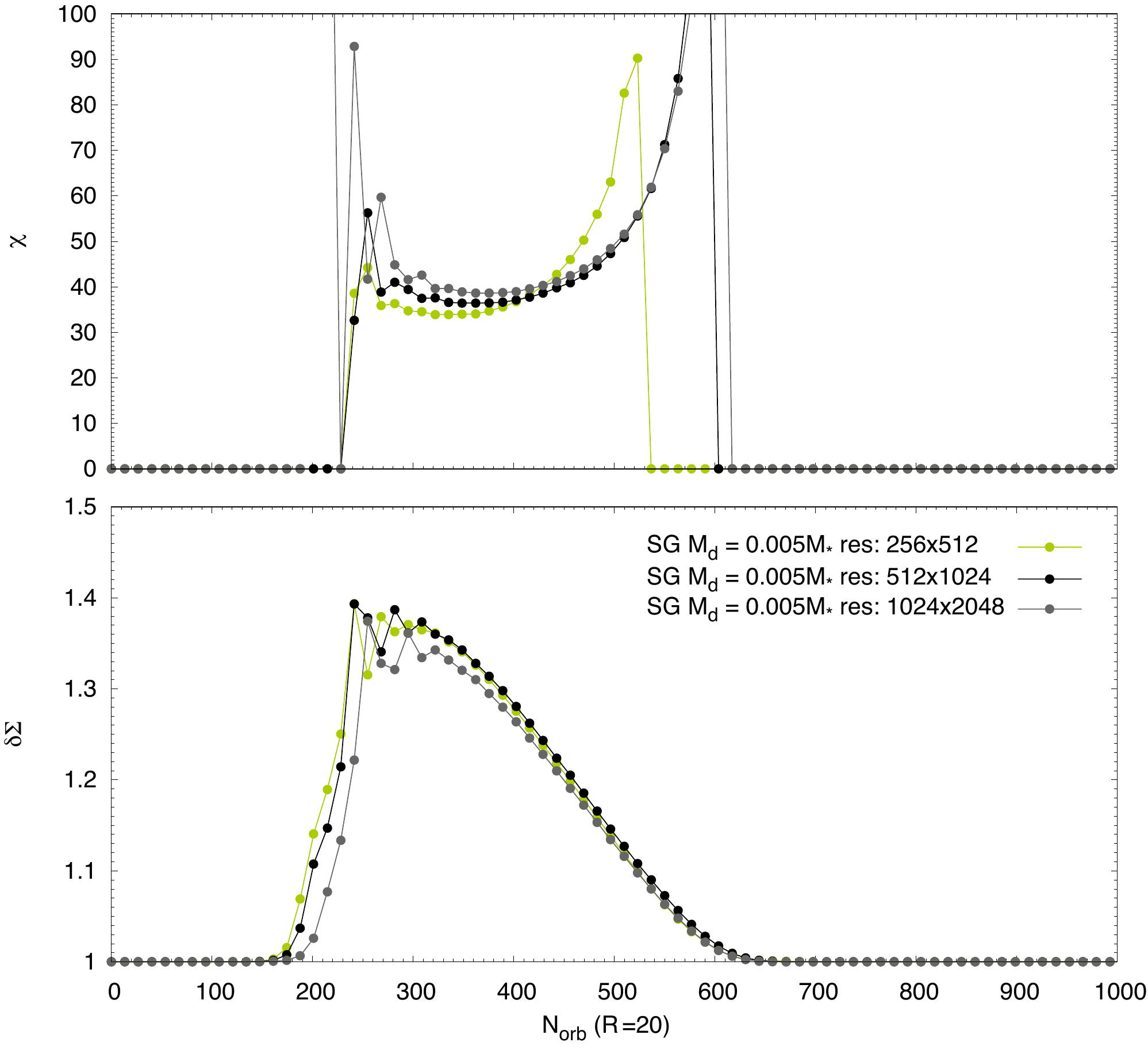}
        \caption{Numerical convergency test for $M_\mathrm{disc}/M_*=0.005$, $\Delta R_\mathrm{dze}=1H_\mathrm{dze}$ and $\alpha=10^{-4}$ model assuming three different numerical resolution. }
        \label{fig8}
\end{figure}

In order to verify the numerical convergency of our simulations, we ran additional simulations with different numerical resolutions. The applied numerical resolutions were: $256\times512, 512\times1024$ and $1024\times2048$. The radial and azimuthal distribution of the grid cells were  logarithmic and equidistant, respectively.  As a result, the disc is resolved by $0.2H,\, 0.1H$ and $0.05H$ everywhere assuming increasing numerical resolution. Fig.\,\ref{fig8} shows the evolution of $\chi$ and $\delta\Sigma$ in the self-gravitating $\alpha_\mathrm{dz}=10^{-4}$ model for $M_\mathrm{disc}/M_*=0.005$ disc. All other parameters of the model are unchanged. 

As one can see, all simulations reveal the same qualitative behaviour of $\delta\Sigma$ and $\chi$. The formation of the full-fledged vortex requires $\sim200$ orbits, independently of the applied numerical resolution. The vortex lifetime is $\sim400$ vortex orbits for the two highest numerical resolutions, and somewhat less, $\sim350$ vortex orbits for the numerical resolution applied throughout the paper. This can be explained by how the vortex radial width and the achieved numerical resolution are commensurable for this particular model. Both $\delta\Sigma_\mathrm{max}\simeq1.4$ and $\chi_\mathrm{min}\simeq40$ agree, independently of the numerical resolution. Generally, we conclude that our simulations using $256\times512$ grid cells are in the numerically convergent regime.

\label{lastpage}
\end{document}